\documentclass[]{elsarticle}

\usepackage{lineno,hyperref,subfig}

\journal{Journal of Theoretical Biology}

\usepackage{xr-hyper}
\usepackage{graphicx}
\usepackage{amsfonts}
\usepackage[version=4]{mhchem}
\usepackage{mathalfa}
\usepackage{psfrag}

\usepackage[margin=1in]{geometry}

\newcommand{\one}{($i$) }
\newcommand{\two}{($ii$) }
\newcommand{\three}{($iii$) }
\newcommand{\four}{($iv$) }

\begin{document}

\begin{frontmatter}
\title{Performance limits and trade-offs in entropy-driven biochemical computers}

\author{Dominique Chu}

\address{ School of Computing \\ University of Kent\\ CT2 7NF, Canterbury\\ 
              Tel.: +123-45-678910\\
              Fax: +123-45-678910         
}
\ead{d.f.chu@kent.ac.uk}
%


\begin{abstract}
It is now widely accepted that biochemical reaction networks can perform computations.  Examples are kinetic proof reading, gene regulation, or signalling networks. For many of these systems it was found that their computational performance  is limited by  a  trade-off between the metabolic cost, the speed and the accuracy of the computation. In order to gain insight into the origins of these trade-offs, we  consider entropy-driven computers as a model of biochemical computation.   Using tools from stochastic thermodynamics, we show that entropy-driven computation is subject to a trade-off between accuracy and metabolic cost, but does not involve time-trade-offs. Time trade-offs appear when it is taken into account that the result of the computation needs to be measured in order to be known. We argue that this measurement process, although usually ignored, is a major contributor to the cost of biochemical computation. 
\end{abstract}


\end{frontmatter}


\section{Introduction }
\label{sec:Introduction}

Computing architectures based on biochemistry, rather than semi-conductor technologies, are attracting increasing interest as alternative models of computation \cite{amos}. Biochemistry can be used to engineer novel types of computers based on biological components.  Examples  include,   DNA based computers  \cite{wersglaubt,lakindna}, robots controlled by slimemolds  \cite{Phi}, or logic gates implemented  in  living cells   \cite{friedenauferden,macia,rochen,milka}. Beside this technological importance of biochemical computers,  there is now also an increasing appreciation that information processing may be an important fitness contributing function for  natural organisms \cite{sarahreview,sarahreview2}. There are a number of biosystems that have been studied as   {\em in vivo} special purpose computations.  For example,    kinetic proofreading  \cite{proofread,dellwo}  greatly enhances the copying fidelity during translation  and is often interpreted as an {\em in vivo} computation.  A classical example of biochemical computation is bacterial sensing   \cite{anfisa,government2,purcell,zubin}, whereby cells measure molecular concentrations in their environment and modify internal pathways  and gene expression levels in response.   Chemotaxis  \cite{noisesurette},  for instance, depends on  organisms sensing a molecular concentration gradient by computing the difference between several measurements, either in time or across the cell volume. Most recently even bacterial growth dynamics has been interpreted as a computational process \cite{myscireppaper,mybmcevobiopaper}. 
\par
Detailed case studies of biological computers often find  performance limits to  biochemical computations. For a  simple gene-switch, Zabet and coworker found  a trade-off between the cost, the accuracy and the speed of the computation \cite{myinterfacepaper,myhonepaper}. Similar trade-offs were established for other biological systems, including  chemotaxis \cite{wlan}, regulation of nutrient uptake \cite{myscireppaper}, and  translation \cite{magna}; lower limits on the cost of sensing (not involving trade-offs) have also been  found recently \cite{keizu,government2}. 
\par
Intuitively such trade-offs are to be expected. Bio-chemical networks are stochastic systems and as such subject to noise. Overcoming this noise requires energy input and time.   
Energy-time-accuracy  trade-offs are also  implied by the classical results on  the physics of computation \cite{landauer,bennetthermo,feynmancomputation}. While there does not seem to be a lower limit for the energy used during a  computation,   Bennett \cite{bennetthermo} pointed out that  in the zero energy limit the speed of computation goes to zero. Computations that complete within a finite time,  therefore require finite energy resources. 
\par
The question is now whether one can go beyond both the individual  case studies of biochemical computers and the intuitive arguments and establish a  model  which provides insights into the origins of the  performance limits to  biochemical computations. The task is a difficult one. For one,  there is a wide variety of approaches to biochemical computation  (only some of which are mentioned above). At the same time,  there is no general definition of biological computation, i.e. it is not clear how to distinguish a reaction network  that computes  from one that does not.
\par
For the purpose of this article, we will take a pragmatic approach with respect to the latter question and simply  identify (in section \ref{define}) computation with out-of-equilibrium biochemical  processes. According to this, every biochemical process that is not in equilibrium performs a computation. As far as the wide variety of biochemical computations are concerned, we will abstract away from specific models and  define  the  concept of    {\em entropy driven computers} (EDC) in section \ref{edc}. This will  capture many  properties of  {\em in vivo} computers as they appear in biological systems.  EDCs  are in many aspects different from real biological networks, but  we  will argue that  they share important characteristics with a wide range of biochemical computers. It is perhaps best to think of EDCs as test-tube biochemistry, as it is frequently used in biological research to study reaction networks {\em in vitro}. 
\par
We model  EDCs as  continuous time Markov chain models of  biochemical systems. We  assume that each  model is  initialised in some state and then left to relax to equilibrium. We will then interpret this  relaxation process  as a computation. Throughout this article we will assume that the EDC  is of mesoscopic scale. By this we mean that it is still affected by stochastic fluctuations, but that it is also within the range of validity of the linear noise approximation \cite{vankamp}. Simply put, this assumption states that the stochastic system behaves like the deterministic equivalent, plus some noise. The linear noise approximation is a very good approximation for mesoscopic systems and  holds true for a wide range of biochemical systems, and hence for a wide range of biological ``computers'' such as gene regulatory networks, protein-protein interactions or intra and inter-cellular signalling systems, although clearly there are  systems that will not be captured by this assumption. 
\par
For the purpose of the present contribution, we will identify the cost in energy of a computation with the entropy produced during the computation.  While this does not quantify the actual metabolic cost of this computation, it is  directly related to it.   We will first show that the linear noise approximation  implies  that the  entropy production scales linearly with the system size, while  the time-scale to approach equilibrium (which we interpret as the computing time) remains invariant.    This means that there is a   trade-off  between the cost of the computation and its accuracy, but there is  no   trade-off involving the speed of the computation. Contrary   to previous work (or apparently so), this suggests that speed-energy trade-offs are not  a fundamental property of biochemical computation {\em per se}
\par
A trade-off involving time emerges only when it is taken into account that the result of the computation must be measured in order for the computation to have any impact in the world. Any measurement of the outcome of the computation in turn  requires a measurement device.  This device needs to  be brought into contact with the computer to determine its state. Device and computer then form a joint system, which initially will be out of equilibrium but  relaxes to an  equilibrium. This relaxation constitutes the measurement process.  Formally a measurement is thus also an entropy driven computation. As we will show below, restoring the computer to its original state, while  at the same time  leaving the measurement device in a state that indicates the result of the computation, requires both energy input and time. It  also leads to a trade-off between the energy used and the speed with which the restoration can be completed with a given confidence. A second trade-off involving time arises from the stochastic nature of the computer. A single measurement only indicates the correct result with a certain probability. Repeated measurements are necessary in order to sample the state of the computer reliably, thus leading to a trade-off between accuracy and time.     

\section{Results}

\subsection{Computation by biochemical systems}
\label{define}
 The current modus operandi in the field of biochemical computing is to identify a  biological system  (such as sensing or proof-reading) as  a computation when it implements  a function that is  naturally interpreted as a computation. This approach enables deep insights into specific examples, but   is likely to miss most instantiations of biochemical computation.  It would be much more  useful to have a concept of biochemical computation that is independent of its function, just as in  computer science computation is defined with respect to a number of specific mathematical models, not by reference to what is computed. 
 \par
 The best known model of computation is   the   {\em Turing machine}. This is a mathematical construct consisting of a ``reading head''  that  is reading and writing  a tape, while  changing its internal states in the process,  until it reaches a   ``halting state,'' at which point the computation stops. It is believed that for every computable function there is a corresponding Turing machine that computes it. Based on this,  one  could  be tempted to  define  a biochemical process as a computation if there is a Turing machine that simulates this process.  This does not work however: The natural equivalent of a halting state in  biochemical systems  is the  equilibrium state, i.e. the  state of the biochemical system where reactions are in detailed balance. Unlike the halting state of a Turing machine, the equilibrium state  is of a statistical nature. This means that on average  there  are no net-fluxes across the network of reactions   \cite{beard1,beard2}. This does not mean, however, that reactions stop. Even in equilibrium there is an ongoing chemical activity.  Crucially though,  the sequence of reaction events is  symmetric in time  \cite{vankamp}, such that an observer would not be able to tell apart  an actual sequence of reactions from a (hypothetical) reversed sequence. Equilibrium is not time-directed.  Computation, on the other hand,    is necessarily time directed,  mapping  a particular input to a particular output.  Equilibrium systems are therefore not able to compute.  Sample paths of equilibrium biochemical systems can still be simulated and are thus computable by Turing machines, whether or not the system is in equilibrium.   This demonstrates  that not all processes that can be simulated by Turing machines are also themselves processing information.  
\par
For the purpose of this paper, we will  adopt  the  simplest  working hypothesis and postulate   that the equilibrium state is the only halting state of biochemical computers.   This implies  that every biochemical system  that is not in equilibrium is in the process of computing. By adopting this definition, we also accept that  most biochemical computers will not do any useful calculations, just as almost all Turing machines do not compute anything of interest.

\subsection{Entropy driven computation}\label{edc}

In this section we define an EDC as a closed, stochastic, biochemical system, denoted by a fraktur S,  $\mathfrak{S}$. The system  does not exchange  particles with the environment. We conceptualise $\mathfrak  S$ as  consisting of a (typically very large) number of discrete microstates  $\mathfrak  s_0, \mathfrak s_1,\ldots, \mathfrak  s_m$ (see SI section 1 for more details).      An EDC  is  initialised in a macrostate $M_0^\mathfrak{S}$ characterised by a specified abundance for each of its  constituent biochemical species at time $t=0$; see SI section 1 for a detailed explanation of what we mean by ``macrostate.'' After a transient period, the biochemical system approaches an equilibrium state  $M_\infty^\mathfrak{S}$ characterised by detailed balance. The approach to equilibrium is the computation.  Strictly speaking, the transition to equilibrium takes an infinite amount of time. In practice, EDCs will be very close to equilibrium after a finite, possibly very short, time.   We will model EDCs here as continuous time Markov chains. Then the  time scale for the approach to equilibrium depends on the kinetic parameters appearing in the master-equation that defines $\mathfrak{S}$.   We will take this  time scale as the {\em speed} of the computation.   
\par
In contrast to EDCs, {\em in vivo} computations, i.e. cells,  never approach equilibrium, but rather operate around non-equilibrium steady states. We acknowledge this, but still argue that for the present purpose  equilibrium models are more convenient. They are also more  revealing of underlying principles, for the following reasons:  Firstly, steady state processes have an additional ``maintenance''  entropy production. This may be substantial, but is not related to the cost of the computation {\em per se}. 
 Mathematically, it is possible to separate the contribution from the relaxation from the contribution to the entropy production that arises from the maintenance of the non-equilibrium state. By focussing on the simpler equilibrium case, we circumvent this mathematical complication that does not add anything to the present purpose.
Secondly, as  will become clear below,  considering equilibrium processes  naturally forces a conceptual separation between the computation {\em per se}  and the reading of the result.  Both are separate processes  that limit computational performance in qualitatively different ways.  Finally, the equilibrium state is conceptually reminiscent of the halting state in Turing machines, which in turn is of fundamental importance for the theory of computation. Indeed, there is an interesting analogy  in the  relationship between on the one hand a non-equilibrium steady state, sustained by an organism and test-tube biochemical systems approaching equilibrium and on the other hand   an operating system --- which is not supposed to halt --- and an algorithm --- which must halt.  Models of entropy driven computation therefore naturally link biology, physics and theoretical computer science. 
With this being said,    the basic conclusions that we will reach     depend primarily on the linear noise approximation, which  in turn depends on system size, not on the distance from equilibrium.

\subsection{Cost of the EDC proper}

\label{cost}
\subsubsection{The model}

We assume that the stochastic system $\mathfrak{S}$ is defined by the master equation, which is a differential equation for the probability to observe  a particular combination of  molecular abundances  $\mathbf n$  at time $t$  \cite{gillespiemaster}.
\begin{equation}
\label{mastereq}
\dot P(\mathbf n,t)= \sum_{i} \left( w_i(\mathbf n- \boldsymbol\sigma_i) P(\mathbf n - \boldsymbol{\sigma}_i,t)  - w_i(\mathbf n) P(\mathbf n,t) \right)  
\end{equation}
where $w_{i}(\mathbf n):= k_i h_i(\mathbf n)$ is  the total rate  of reaction $i$ and $h$ is the multiplier indicating how many combinations of molecules can realise this reaction; $\mathbf n$ and $\boldsymbol\sigma_i$ are the particle vector  and the stoichiometric vector of reaction $i$. Altogether, the right hand side of the equation contains two terms. The first one  expresses the possibility that the current state characterised by a particle composition  $\mathbf n$ was obtained from reaction $i$ and the state before the reaction happened was   $\mathbf n - \boldsymbol{\sigma}_i$. The second term formulates the possibility that before the last reaction the state of the system was $\mathbf n$, but  reaction $i$ took it away from this state (to a new state $\mathbf n + \boldsymbol{\sigma}_i$).

\par
 Given such a stochastic model, we  can scale the total number of particles in the initial conditions by a factor $c$ and the  reaction rate constants of bimolecular reactions by $1/c$. This amounts to scaling the volume $V$ of the system while keeping the concentration fixed.  In this way we construct an equivalence class $\mathcal S$  of systems $\mathfrak{S}$. In general, the members  in this class will behave differently. Most importantly, they show different amounts of fluctuations and  approach equilibrium at different speeds.   For mesoscopic   volumes, however, the behaviour of the master equation is increasingly well approximated by the  first order  linear noise approximation  \cite{vankamp} whereby the mean behaviour of the system is invariant to scaling.  Scaling the system size  in the regime of the linear noise approximation  only affects the noise around the mean behaviour, which scales with $V^{-1/2}$. Importantly, scaling the volume does not  affect the time-scale to reach equilibrium, which is determined by the mean behaviour.   In the limiting case of an infinite volume the noise goes to zero and  the system is described by a differential equation whose trajectory corresponds (for monostable systems)  to the mean of the linear noise approximation. 

\subsubsection{Entropy production}

In this sub-section, we will  show that the operation of an EDC is subject to a trade-off between the noise and the cost. An EDC  starts with the stochastic  system $\mathfrak{S}$ initialised in some macrostate  $M_0$. The system then  relaxes into an equilibrium state  $M_\infty$, performing a computation in the process. We will identify the cost of the computation with    the  amount of  entropy  generated during the relaxation.  There are two components to the entropy. Firstly, the  exported entropy or  heat dissipated to the environment  which accrues whenever the system makes the transition from state $\mathfrak s_i$ to state $\mathfrak  s_j$. This is given by the ratio of the forward rate of the transition $w_{ji}$ and the corresponding backwards rate $w_{ji}$  \cite{seifertentropy}. 
\begin{equation}
\Delta S_Q= k_B\ln \frac{w_{ji}}{w_{ij}}
\end{equation}
Using this definition the heat dissipated to the environment is positive, whereas heat extracted from the reservoir is negative.  No heat  is  generated by a system that transitions from a state $\mathfrak  s_i$ to a state $\mathfrak  s_j$ and then back again. More generally  the heat generated by a system that  transitions from an initial state $\mathfrak  s_0$  to some final state  $\mathfrak  s_j$ via a number of intermediate states is independent of the particular path, and depends only on the initial and final state. Taking into account that the heat generated in each transition is additive, and applying the detailed balance condition that relates the steady state probabilities  $\pi_i, \pi_j$ of states $\mathfrak  s_i$ and $\mathfrak  s_j$ of a stochastic system with the transition probabilities, i.e. 
\begin{equation}
\pi_i w_{ij} = \pi_j w_{ji} \tag{detailed balance condition}
\end{equation}
we  obtain:
\begin{equation}
\label{ente}
\Delta S_Q(\mathfrak  s_0 \rightarrow \mathfrak  s_j)= k_B\ln \frac{\pi_j}{\pi_0}
\end{equation}
\par
In addition to the exported entropy, the computation also produces Shannon entropy. This  Shannon entropy of a microstate $\mathfrak s_i$ can be written as $\Delta S_\mathrm{int}= -k_B\ln p_i$, where $p_i$ is the probability to observe the system $\mathfrak{S}$ in microstate $\mathfrak s_i$. On average this gives  a contribution $\langle S_\mathrm{int}\rangle= -k_B\sum_i p_i \ln (p_i)$. The Shannon entropy  produced during the computation is then  given by $\langle S_\mathrm{int}(\infty)\rangle - \langle S_\mathrm{int}(0)\rangle$. 
%
\begin{figure*}
\centering\includegraphics[angle=-90,width=0.45\textwidth]{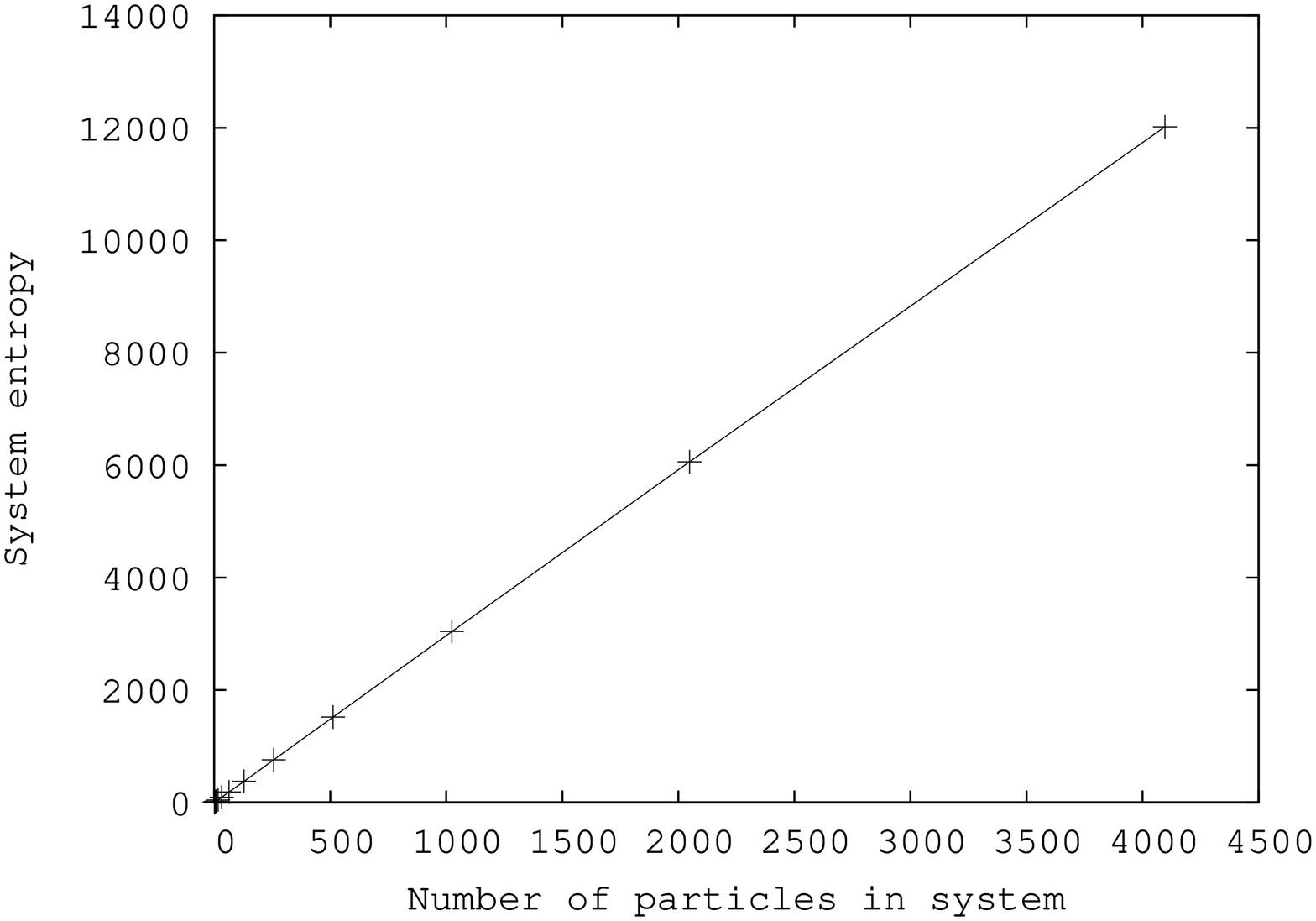}
\centering\includegraphics[angle=-90,width=0.45\textwidth]{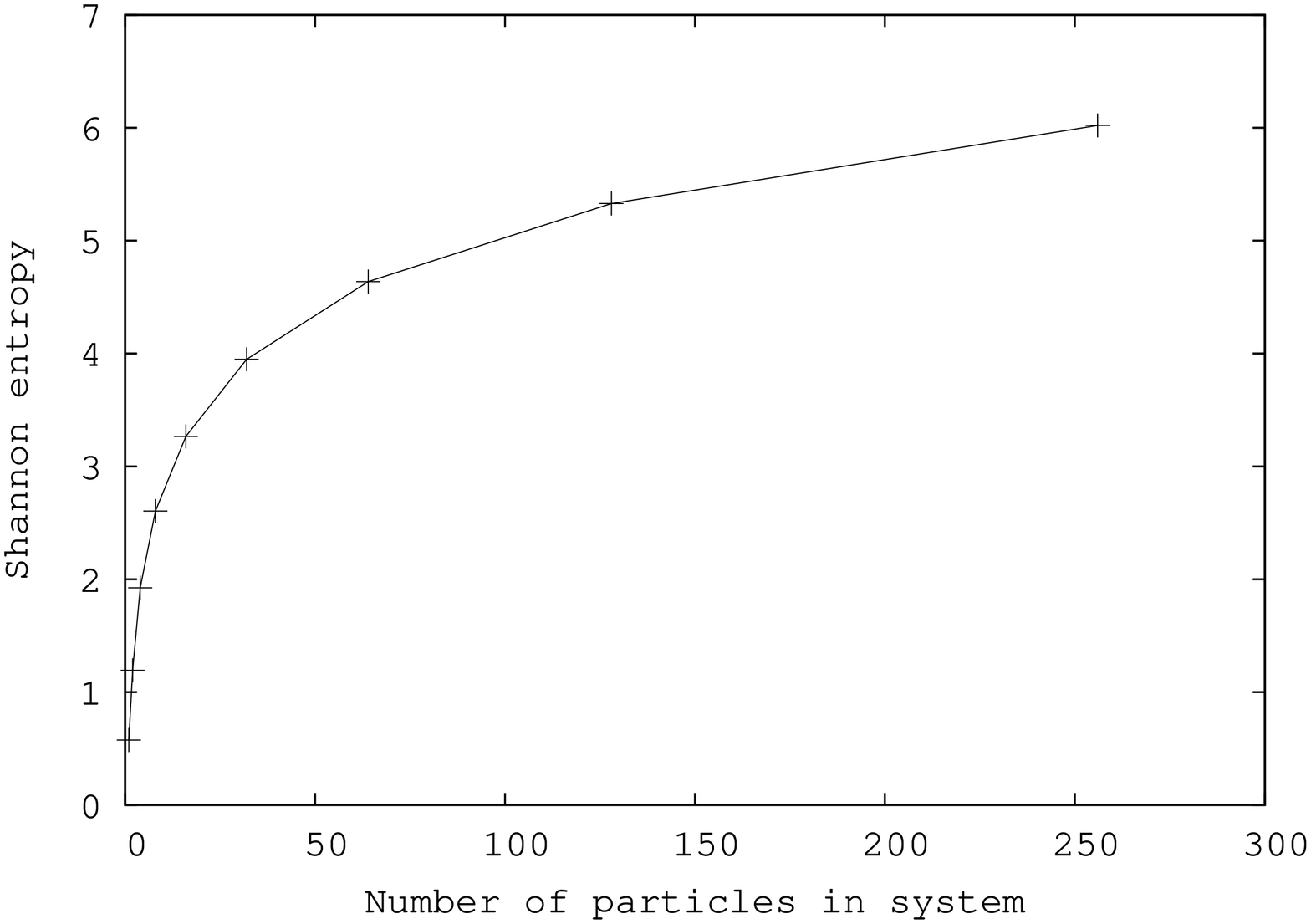}
\caption{
Numerical calculations of the exported entropy  ({\bf left}) and the Shannon entropy ({\bf right})  for the system $A + B \rightleftharpoons C+ D \rightleftharpoons E$. The system size is the number of  $A$ and $B$ at time $t=0$. The concentration was kept constant as the particle size was increased.   The entropy  is given in units where  $k_B=1$. The forward rate constant  was set to $0.3$ and the backwards rate constants from $E$ and $C+D$ were $0.05$ and $0.1$ respectively.  Prism \cite{prism} was used to solve the numerics; for the model file see supplementary information.}\label{exampleres}
\end{figure*}
\par
For realistically sized systems it is usually not possible to solve the master eq.~\ref{mastereq} exactly, even in steady state. Analytic expressions for the entropy production are therefore difficult to obtain. However, general scaling arguments can be given in the limit of the linear noise approximation. In this regime, one expects the total  entropy  during the approach to equilibrium  to scale linearly with the system size.  Consider, for example, a biochemical system with a mono-modal steady state distribution and discrete, steady state probabilities $\{\pi_0,\pi_1,\ldots,\pi_k,\ldots,\pi_m\}$, where  $m + 1\gg 1$ is the total number of states of the system. Assume now that  the initial state is  $\mathfrak  s_0$, then the heat dissipation is given by averaging eq.~\ref{ente} over all possible end states:
\begin{equation}
\label{heat}
\langle S_Q\rangle = k_B\sum_{i=0}^m \pi_i\ln\left({\pi_i\over \pi_0}\right)=k_B\sum_{i=0}^m\pi_i\ln\left({\pi_i}\right) -k_B\ln{\pi_0}.
\end{equation}
The first term on the right hand side equals   the Shannon entropy (up to the sign).  The total entropy produced  $\langle S\rangle= \langle S_Q\rangle + \langle S_\mathrm{int}\rangle$ is therefore given by 
\begin{equation}
\langle S\rangle = -k_B\ln \pi_0.
\end{equation}
In order to understand the scaling of the entropy production, we approximate the discrete steady state distribution by a continuous Gaussian distribution with mean $\mu$ and variance $\sigma^2$.   According to the linear noise approximation, the variance and the mean scale linearly with $V$, such that we obtain:
\begin{equation}
\langle S\rangle = -k_B\ln \pi_0 \sim {x_0^2 - 2x_0\mu + \mu^2 \over 2 \sigma^2 }\sim V.
\end{equation}
\par
We stress that this scaling argument is only valid for mesoscopic systems when the linear noise approximation is good.  In particular, for very small systems  there can be a non-linear relationship between the system size and the  heat dissipation. In those cases the master equation needs to be solved to determine the entropy production.  This  is usually problematic.  See SI section 5 for an explicit calculation of the heat dissipation for a minimal example and figure \ref{exampleres} for a graphical representation.  
\par
In summary: The ability to determine the result of the EDC is limited by the noise of the system in equilibrium, which  scales like  ${V}^{-{1\over 2}}$ in the linear noise approximation.  The cost of the computation scales linearly with the size of the system.  Hence there is a trade-off between entropy production (and thus energy cost) and the accuracy of the EDC.  No time-trade-offs arise here. In the linear noise approximation, the time evolution of the mean does not depend on the system size.

\subsection{Trade-offs arising from the measurement}
\label{measurement}
The equilibration of the EDC is only one part of the computation. In order, for the outcome of the computation to be known, a measurement must be performed to determine  the macrostate  $M_\infty^\mathfrak{S}$ and the  result needs to be recorded. This comes  at a cost \cite{eule,perscopy}.   We assume that the measurement is done by means of the measurement device $\mathfrak{\bar S}$, which is itself an EDC; see SI section 1 for details on the concept of measurement. 
\par
We will  first discuss how  $\mathfrak{\bar S}$ can  be used for  binary measurements on observables of $\mathfrak{S}$, i.e.~in order to determine whether a particular biochemical  species $L$ of $\mathfrak{S}$ is above or below a threshold abundance. We will assume that  $\mathfrak{\bar S}$ is a bistable system, i.e.~in equilibrium, it can be in one of two (transient) macrostates, indicating the result of the computation.   
\par
Measurement is only possible if $\mathfrak{S}$ and $\mathfrak{\bar S}$ are temporarily brought into contact.  We will  assume that the contact  remains limited to specific interfaces. The biochemical systems  $\mathfrak{S}$ and $\mathfrak{\bar S}$ should not be allowed to   mix, as it would be difficult to separate the systems again  after the measurement.   Instead, we consider here a protocol whereby  $\mathfrak{S}$ and $\mathfrak{\bar S}$ remain separated by a wall. We model the interface between system and measuring device as a single ``trans-membrane'' receptor placed in the surface of   $\mathfrak{\bar S}$. The external part of the sensor contains a number of binding sites for molecules of type  $L$  of  $\mathfrak{S}$. The inside of the receptor can interact with $\mathfrak{\bar S}$, as will be described in the next section.  We will assume a  Monod-Wyman-Changeux (MWC)  \cite{marzen} receptor (but other models are possible too).  When its external  sites are bound then the receptor  is heavily biased towards the ``active'' state. Otherwise it is heavily biased towards being deactivated.  When there are several binding sites, then the receptor can support   ultra-sensitivity. This means that for low abundances of $L$ the binding sites are almost never bound; above a threshold abundance the sites are almost always bound. In this way, MWC receptors can be used for binary measurements of external concentrations. 
\par
Upon contact, the system and the measurement device form a new system  $\mathfrak{S}\mathfrak{\bar S}$. The joint system will therefore evolve from an initial macrostate   $M^{\mathfrak S\mathfrak{\bar S}}_t$ at time $t$ of the measurement,  to a macrostate  $M^{\mathfrak S\mathfrak{\bar S}}_{t+ T_\textrm{meas}}$, where $T_\textrm{meas}$ is the time required for the measurement. Upon separation of  $\mathfrak{S}$ and $\mathfrak{\bar S}$  the new macrostate of the measurement device should be  a (transient) record of the state of $\mathfrak{S}$. Once the measurement is completed it is necessary to restore $\mathfrak{S}$ as an independent system, i.e. to separate $\mathfrak{S}$ and $\mathfrak{\bar S}$. 
\par
Conceptually, there are thus two aspects to the measurement process that are relevant for the question we consider here. Firstly, the process of measurement itself and secondly, the  separation of the measurement device from the system $\mathfrak{S}$. We find that the binary measurement ---  while it cannot be done for free --- does not imply any trade-offs. In contrast, the separation step leads to a trade-off between energy, cost and time.   
\par

\subsubsection{The measurement proper}

We choose   $\mathfrak{\bar S}$  to be  bistable, i.e.~there are two  macrostates $M^m$ and $M^g$ characterised by a high amount of $m$ or $g$ respectively.  Bistability can  occur in equilibrium biochemical systems. Stochastic bistable systems will switch spontaneously, but possibly rarely,  between the two macrostates. The expected time between switching events will depend on the system size and the  ``well depth.'' The latter is essentially determined by the kinetic parameters of the system and indicates the difficulty of escaping a metastable state. Larger systems are more stable but even  relatively small  systems  can be  sufficient to virtually guarantee stability over any  practically relevant time-scale. 
\par
  In order to switch a bistable  system at  a {\em particular time}, it is necessary to take it out of equilibrium by introducing a net probability flux to the desired state of the system. This implies breaking detailed balance and hence dissipates heat. A controlled switch of the bistable system can therefore only be achieved   at a certain expense of work.   Formally, it would be sufficient to connect a source of free energy (a ``chemical battery'') to  $\mathfrak{\bar S}$ during the switch only  and disconnect it afterwards.  In practice, due to the nature of the biochemical systems, it is difficult to remove such a battery without extra work input.  It is therefore much better to operate the device permanently from a battery away from equilibrium.

\subsubsection{A possible design of a measurement device}

We will now construct a measurement device    $\mathfrak{\bar S}$ consisting of  the molecular species $\{c,d,e,g,m\}$. It is   bistable   in $m$ and $g$, i.e. at any one time only one of those two species is present in a high concentration. We will identify the two metastable states by the macrostate $M^g$ and $M^m$ respectively. These states  will indicate whether  the  concentration of species $L$ of $\mathfrak S$ is above or below some threshold. There are many ways to achieve bi-stability in biochemical systems. We choose a mechanisms based on two competing auto-catalytic reactions:  Molecules of type  $c$  are converted into $d$. This conversion is  catalysed by $g$. Molecules of type  $d$  are reversibly converted into $g$. Altogether, $g$ is thus autocatalytic. To achieve bi-stability, we add symmetrically a second autocatalytic circuit consisting of $e$ which is produced from $c$ catalysed by $m$. Molecules of $e$ are also reversibly converted into $m$. The reactions can be summarised as follows: 
%
\begin{eqnarray*}
&&\ce{$c$  <=>[$g$] $d$}\\
&&\ce{$c$  <=>[$m$] $e$}\\
&&\ce{$d$ + $d$ <=> $g$}\\
&&\ce{$e$ + $e$ <=> $m$}\\
\end{eqnarray*}
To create bi-stability, this system needs to be extended  by an antagonistic interaction between $g$ and $m$. 
\begin{eqnarray*}
\ce{$d$ + $m$ <=> $m$ +$c$}\\
\ce{$e$ + $g$ <=> $g$ + $c$}\\
\end{eqnarray*}
Finally, we also posit that $g$ and $m$ inter-convert reversibly, albeit at a very low rate. Forward and backwards rates are equal. 
\begin{displaymath}
\ce{$g$ <=> $m$ }
\end{displaymath}
If initialised with a sufficient number of $c$ and one $g$ and $m$ each, then  the system will, after a transient period, evolve to a macrostate $M^g_\infty$ where  $g> m$ or $M^m_\infty$ characterised by $m>g$. The probabilities of either of these states will be equal if the parameters are symmetric between the autocatalytic pathways of $m$ and $g$.  
\par

\subsubsection{Switching}

%
\begin{figure}
\centering\includegraphics[angle=-90,width=0.45\textwidth]{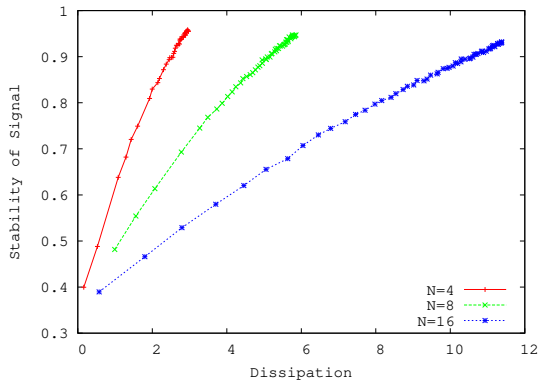}
\caption{The trade-off between dissipation and stability. For networks of different size $N$, defined as the total number of particles in the system,  the energy dissipation after 600 time units was recorded against the stability of the memory. Small systems are performing better than large systems. Stability is defined as the number of $m$ molecules divided by  the number of $m$ and $g$ molecules combined after 600 time units. Each point in the graph represents the average over 5000 individual simulations. For each value of $N$ we sampled initial amounts of ATP from 5 to 300 in steps of 5. The dissipation is the difference between the number of ATP at time 0 and at time 600 averaged over 5000 simulations.} 
\end{figure}
In order to switch  the state of $\mathfrak{\bar S}$, we couple the  chemistry of the device   to the internal part of the MWC receptor, which can be in an active state $r^*$ or an inactive state $r$. Here we assume that the active receptor catalyses the inter-conversion of  $m$ and $g$, adding the following reactions to the measurement device:
\begin{eqnarray*}
\ce{r <=> r$^*$}\\
\ce{m  <=>[r$^*$] g}
\end{eqnarray*}
In the absence of a  driving force, the active receptor  accelerates the approach to $m=g$. Given the right range of parameters it will thus temporarily disrupt bi-stability and bring $\mathfrak{\bar S}$ into a mono-stable regime. Upon removal of the catalysing reaction, i.e. when the internal receptor reverts to the inactive state, $\mathfrak{\bar S}$ will relax quickly to $M^g$ or $M^m$, both with equal probability. No directed switch is possible.
\par
A directed switch is only possible if the catalytic reaction is driven preferentially into one direction. One possibility  is to  couple it to a biochemical battery, i.e. a reaction that is not in equilibrium.  A modified reaction scheme could then be the following:
\begin{displaymath}
\ce{$m$ + ATP  <=>[$r^*$] $g$ + ADP}
\end{displaymath}
If there is a large excess of ATP over ADP then this would result in a net drive towards $g$ whenever the  receptor is activated. Assuming the excess of $g$ is sufficient, the activation of the receptor leads to a switch of the macrostate  to $M^g$.

\subsubsection{Measurement and recording}

The  measurement procedure is as follows: \one  Reset $\mathfrak{\bar S}$.  \two  Initiate the   measurement proper on $\mathfrak{S}$ by bringing   $\mathfrak{S}$ and $\mathfrak{\bar S}$ into contact mediated by the MWC receptor.  \three Wait for a fixed amount of  time $T_\textrm{meas}$. \four  Separate   $\mathfrak{S}$ and $\mathfrak{\bar S}$. 
\par
The reset step is necessary so as to ensure that the measurement device is in a known state prior to starting the measurement. Without this step, $\mathfrak{\bar S}$ is in state $\bar M^m$ with probability $1/2$. This is problematic, because the  measurement of $\mathfrak S$ will only effect a switch in $\mathfrak{\bar S}$ if the concentration of $L$ is indeed above the threshold. The reset can be implemented  as a measurement of  some species $I$ of a  reference volume that is in a known state.  The measurement should be mediated by an auxiliary receptor different  from the one that measures $\mathfrak S$.  The active state of this auxiliary receptor should effect a switch to $M^g$ when the concentration of  $I$ in the reference system is high (which should always be the case).
\par
Once the reset is completed, the measurement of $\mathfrak S$ can be performed. Following this step  the macrostate  of $\mathfrak{\bar S}$ is a record of the state of $\mathfrak S$. The metastable states of $\mathfrak{\bar S}$ are only stable over a finite (although possibly very long) time, because the system may transition spontaneously to a different state. 
\par
In principle it is possible to remove the battery from the bistable system  post-measurement. In practice this will be difficult because it entails extracting the ATP and ADP molecules from  $\mathfrak{\bar S}$ and  requires biochemical work. Therefore, it is better not to remove the battery. In this case, however,  the macrostate of $\mathfrak{\bar S}$  remains only stable  for as long as the battery is sufficiently charged, i.e. as long as  there is an  excess of ATP over ADP. Once the battery has run out $\mathfrak{\bar S}$ will no longer be bistable (see SI fig.~1).
\par
  The  biochemical battery is discharged by two processes. Firstly, during switching the receptor is active with high probability and  the battery will drive the conversion from $g$ to $m$ while using ATP. When the external binding sites of the MWC receptor are occupied the receptor will be active most of the time, resulting in a high rate of discharge. Secondly, there is spontaneous activation of the receptor even if no ligands $L$ are binding to the outside receptor. The rate of spontaneous activation of the receptor  determines the  base-rate of ATP usage/battery discharge.
\par
Note that the ability of $\mathfrak{\bar S}$ to measure and record the state of $\mathfrak S$ is not subject to a trade-off between  system size and accuracy. In the particular model we present here, small measurement devices $\mathfrak{\bar S}$ perform better than those with a large number of molecules, while also dissipating substantially less energy (see SI fig.~1). Similarly, there are no trade-offs between the speed of the measurement and its accuracy. The speed is related to the scale of the reaction rates  inside  of $\mathfrak{\bar S}$ which does not affect the heat dissipation of the system.  
\par
There are, however, other sources of trade-offs that are closely connected to the measurement process.  The accuracy of the binary measurement corresponds to the ability of  $\mathfrak{\bar S}$ to  indicate whether or not the concentration of the system is below or above a threshold.  The parameter that determines this accuracy is the number of binding sites for the ligand $L$.   The receptor number does not influence the cost of switching, but it does increase the cost and time of separating $\mathfrak{S}$ and $\mathfrak{\bar S}$, as will be discussed in the next section.

\subsubsection{The separation step} 

Once the measurement of the system has been completed the coupled system $\mathfrak{S}\mathfrak{\bar S}$ needs to be separated, while restoring $\mathfrak{S}$ to  a state that  is  statistically equivalent to the state  before the measurement. This simply  reflects the condition that the measurement should not alter the system. In our specific case the restoration of the system requires  the separation of $\mathfrak S$ and $\mathfrak{\bar S}$ and that the ligand $L$ is returned to the system. We will find that this separation step gives rise to a trade-off involving time.    To do this, we use an abstract model  of separation which, using ideas from thermodynamics, estimates minimal work requirements. While the model is not biological realistic, we will nonetheless find that  it leads to biologically relevant trade-offs  in terms of binding rates that also  relate to known limitations of biochemical sensors. 
\par
\begin{figure}
\psfrag{x1}[][b]{$x_1$}
\psfrag{x0}[][b]{$x_0$}
\psfrag{x2}[][b]{$x_2$}
\psfrag{x3}[][b]{$x_3$}
\psfrag{s}[][b]{$\mathfrak{S}$}
\psfrag{sb}[][b]{$\mathfrak{\bar S}$}
\psfrag{v1}[][b]{$V_1$}
\psfrag{v2}[][b]{$V_2$}
\psfrag{(i)}[][b]{\one}
\psfrag{(ii)}[][b]{\two}
\psfrag{(iii)}[][b]{\three}
\psfrag{(iv)}[][b]{\four}
\centering\includegraphics[width=0.9\textwidth]{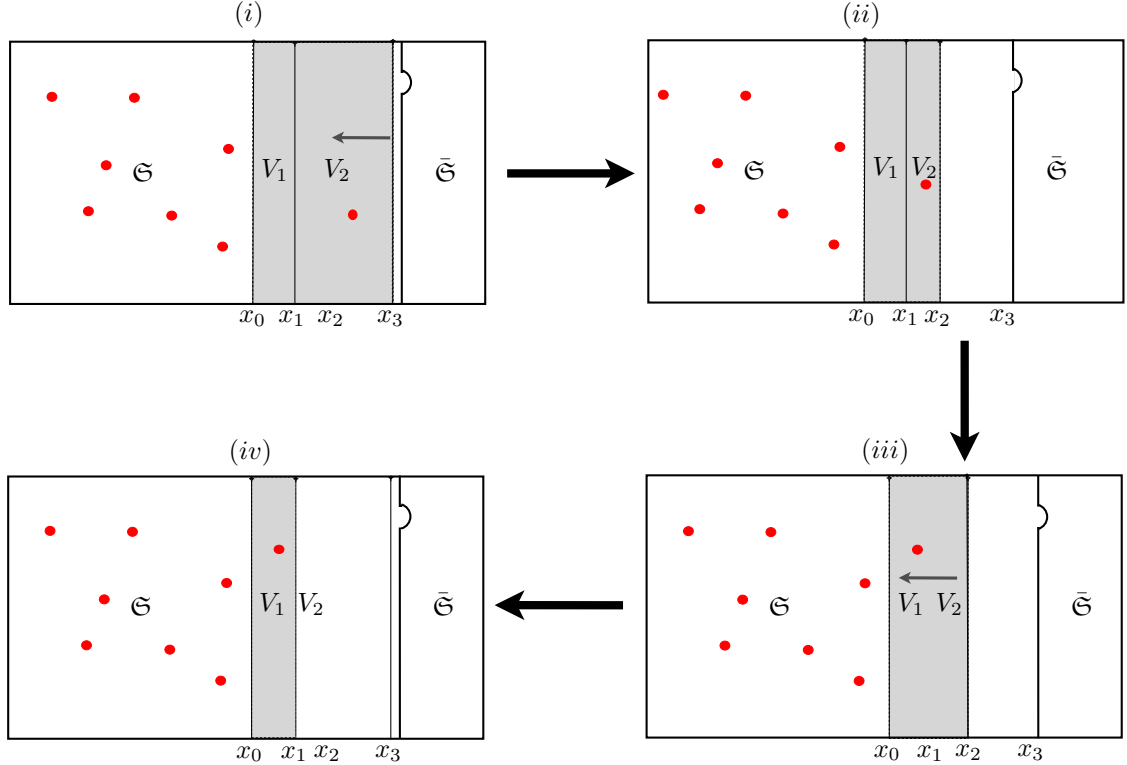}
\caption{A schematic representation of the resetting procedure. The steps \one to \four are illustrated from top left to bottom right. In this particular case, the particle was successfully captured in the first step and is thus successively transported into the volume $V_1$. }\label{szilardreset}
\end{figure}

  We  stipulate that  contact between $\mathfrak{S}$ and $\mathfrak{\bar S}$ is mediated by a straight wall. The system $\mathfrak{S}$ is fixed in space, but   $\mathfrak{\bar S}$ can be moved to the right, which  opens up a volume $V_0$ between the two systems (see fig. \ref{szilardreset}). We also assume a number of  removable walls  that  can be inserted at arbitrary (but  fixed)  points to further sub-partition $V_0$. We now describe a  protocol to restore $\mathfrak{S}$. This model involves 3 walls. A simpler --- but energetically less favourable --- model is possible and described in the SI section 4.  

The idea is to insert a  wall (Wall 1)  close to the membrane of $\mathfrak{\bar S}$ at $x_0$; see fig.~\ref{szilardreset}. It separates $\mathfrak{S}$ from $\mathfrak{\bar S}$. The measurement device is then moved to the right,  opening up the volume $V_0$.  Wall 2 is then inserted  at some  distance to the right of  wall 1 at position $x_1$ so as to  form the volume $V_1$ between it and wall 1.  The separation procedure is as follows. \one After waiting $T$ time units  insert  wall 3  at $x_3$, which is immediately to the left of the membrane of $\mathfrak{\bar S}$. Wall 3 separates the receptor (and possibly ligands bound to it) from the volume $V_0$.\two  Slide wall 3  to the point $x_2$. Wall 3 now  forms a volume $V_2$ between itself  and  wall 2 and a volume $V_0-V_1-V_2$ between itself and the membrane of $\mathfrak{\bar S}$. \three  Remove  wall 2; this opens the volume $V_1+V_2$ between $\mathfrak{S}$ and  wall 3. \four Slide wall 3   from position $x_2$ to position $x_1$. At this point the cycle is started again at \one by inserting  wall 2 at $x_3$.   \par
The minimal work required to separate the system is given by (see SI section 3 for the details of the calculation):
\begin{eqnarray}
W_\mathrm{tot}&=&\ln  \left( \frac {x^{-1}}{\kappa+x^{-1}} \right) + \nonumber\\
&+& \ln\left( V_2 \left( x^{-1}+\kappa \right) \over \left( V_0-x^{-1}- \kappa \right) \left(x^{-1}+\kappa+V_2 \right) \right) -1
\end{eqnarray}
Here $0\leq  \kappa\leq V_0-V_2 $ is a constant expressing the size of $V_1$; see SI section 3 for details.
This shows that  if the  concentration    $x\gg 1$ of particles in $\mathfrak{\bar S}$  is high, then    the work scales as the logarithm of the concentration. More importantly though, within   any particular equivalence class $\mathcal S$ of systems, where the concentration does not change, the separation work is independent of the system size and not affected by scaling of the volume $V$. 

\subsection{Trade-offs involving time}

\begin{figure*}
\psfrag{teta}{\huge $\tau \langle n_\epsilon\rangle$}
\psfrag{Time}{\huge $\tau$}
\subfloat[]{\includegraphics[angle=-90,width=0.45\textwidth]{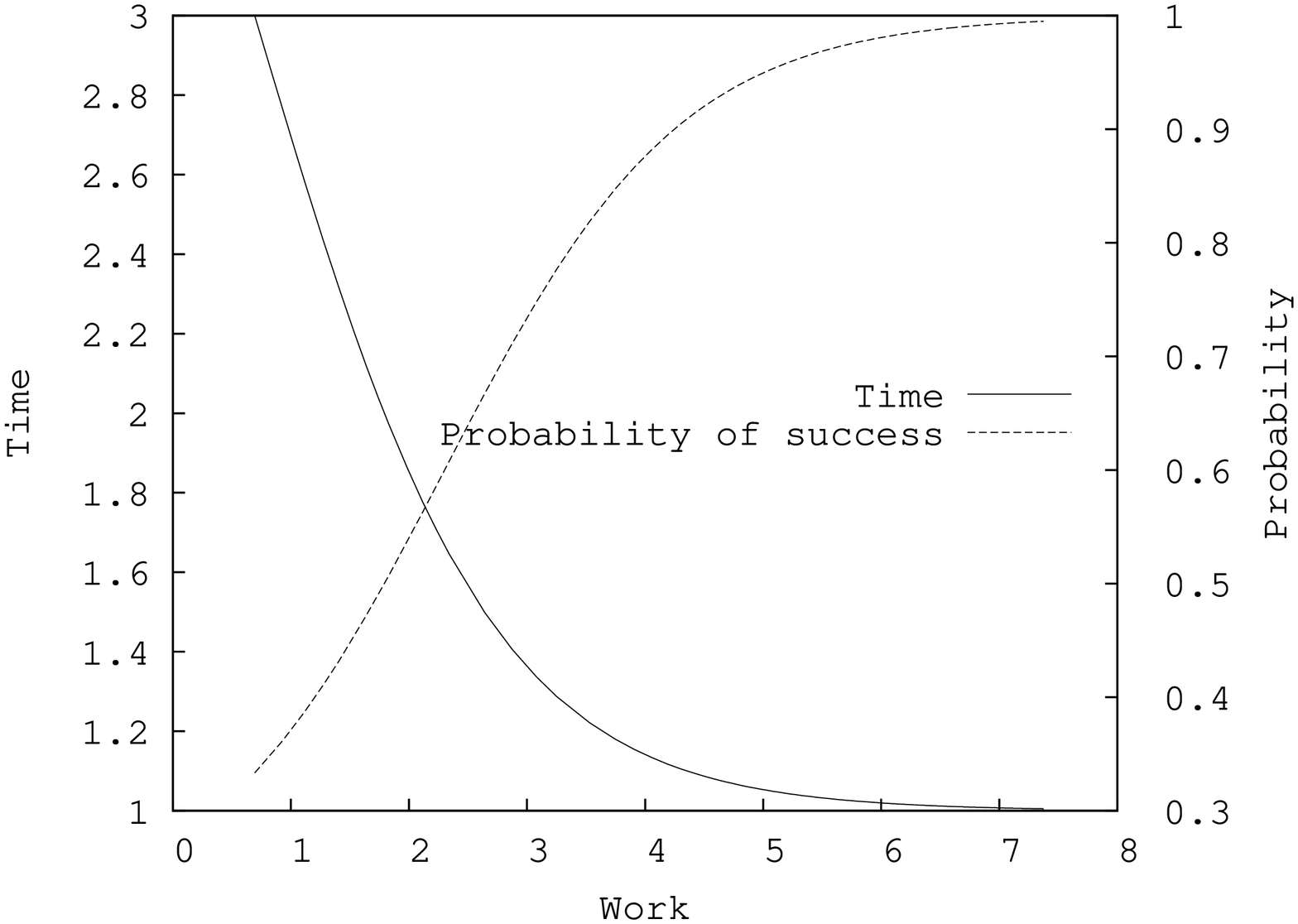} \label{szilardwall}}
\subfloat[]{\includegraphics[angle=-90,width=0.45\textwidth]{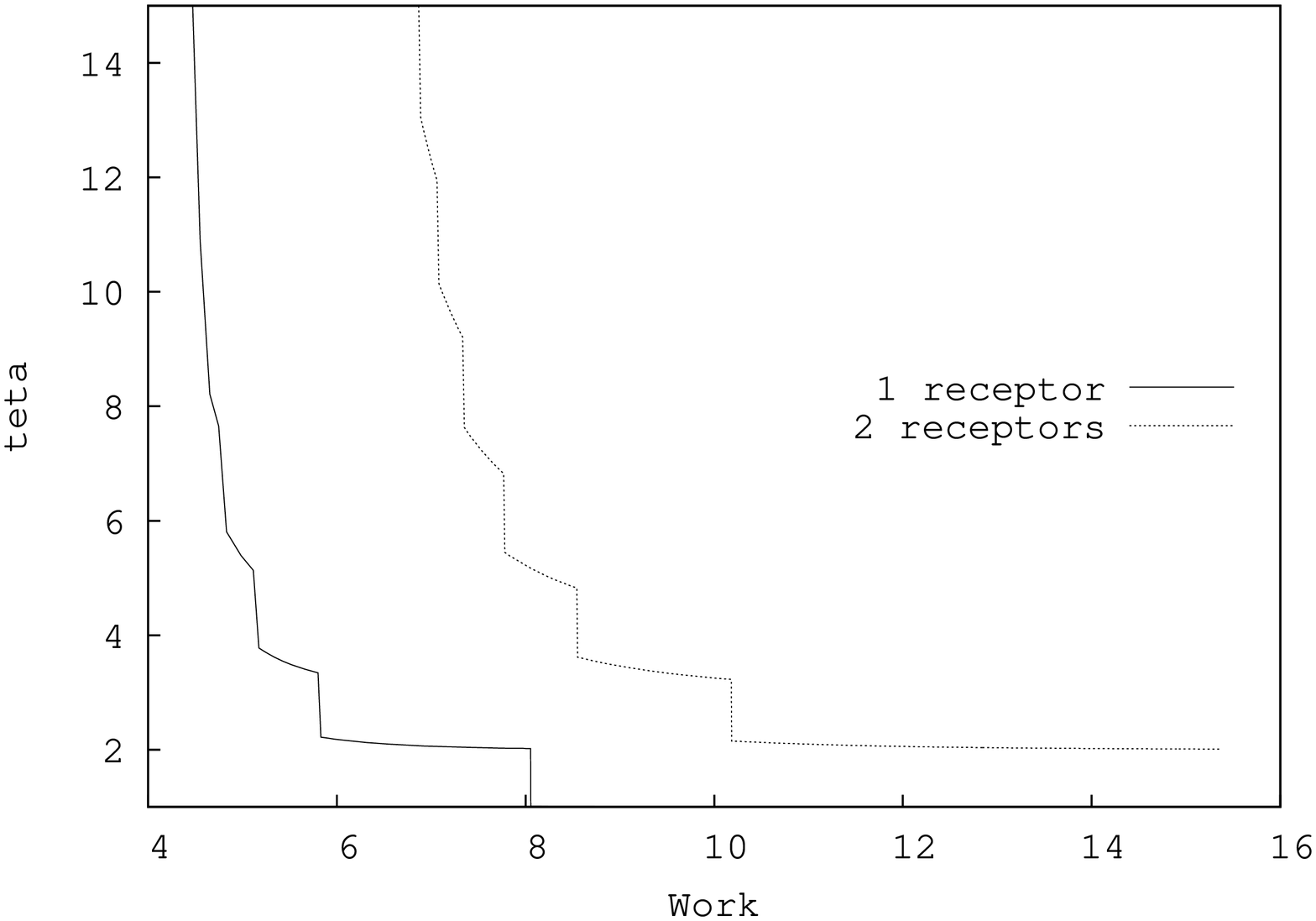} \label{timeworktradeoff}}
\caption{ \protect\subref{szilardwall} Trade-off between the time and the work spent on  the restoration. Here we chose $k=k_u=1$ and $V_1=V_2=0.25$. The reaction volume $V_0$ was varied from $0.5$ to 200.  The  line labelled "Probability" denotes the probability that the particle bound to the receptor has been ``captured,'' i.e. pushed to $V_1$. \protect\subref{timeworktradeoff} Trade-off between the time spent on computation and the work, for 1 and 2 receptors. Here we chose $\epsilon =0.01$, $k/k_u=1, N=200, V_\mathfrak{S}=10$ and $V_2=1$.}
\end{figure*}

The separation step of the two systems gives rise to two different trade-offs involving time. Firstly, the work estimates calculated above are valid for quasi-static processes. Finite time processes will always require more work  \cite{jepsen}. A second trade-off emerges from  the  waiting time for the ligand-receptor bond to equilibrate during step \one of the separation procedure. The time scale  for equilibration is  $\tau\sim k_u + k_b/V_0$, where $k_b$ and $k_u$ are the binding and binding/dissociation rates to/from the receptor. The equilibrium probability  to find the particle unbound is  $p_u= V_0/(V_0 + \eta)$ where $\eta= k_b/k_u$. From this, we  obtain   the number of iterations $\langle n_\epsilon\rangle$ of  the resetting procedure  necessary to ensure  that the average number  of bound ligands is $\epsilon$:
\begin{equation}
 \langle n_\epsilon\rangle= \left\lceil{\ln\left({\epsilon\over l}\right) \over \ln (\eta) + \ln{V_0 + \eta} }\right\rceil,
\end{equation}
where $\lceil y\rceil$ denotes the smallest integer $> y$ and $l$ is the number of ligand binding sites at the external receptor. Multiplying this by  the time scale $\tau$ gives a time-scale indicator for the restoration of $\mathfrak{S}$ (see fig. \ref{timeworktradeoff}).  A trade-off between the measurement time and the work required arises here, because both quantities depend on the volume $V_0$. A larger volume makes it more likely for the unbinding receptor to be captured, but equally increases the amount of work that needs to be done (see fig. \ref{szilardwall}).  
\par
Further trade-offs arise when a concentration needs to be determined with an accuracy higher than binary. To see this, assume  that  $d_\textrm{max}$ is the maximal  abundance of $L$ in $\mathfrak{S}$. Given a set of perfect binary  measurement devices $\mathfrak{\bar S}_i$ one could use half-interval search to determine the abundance of the particles with an accuracy of  $d_\textrm{max}/N$  by using  $\lceil\log(N)\rceil$ sequential  binary measurements.  Each binary measurement comes at a fixed cost and  takes a finite amount of time. Altogether, this means that a higher accuracy can only be achieved at the expense of a longer waiting time and a higher energy cost.
Instead of performing half-interval search, one could alternatively perform  a number of concurrent measurements. In this case, there would be  no additional time penalty, but $N$ measurements would be required in order to achieve an accuracy of $d_\textrm{max}/N$, which is much worse than the logarithmic scaling of the half-interval search.
%

\section{Discussion}

Computational processes in biological systems typically show a trade-off between cost, accuracy and speed. Intuitively such trade-offs are expected, but their precise origin  remained unclear. Our model shows that  the origins of time trade-offs and trade-offs involving accuracy are quite distinct.  The key-parameter controlling the accuracy of the computation of an EDC is  the system size, which also controls the cost.  Yet, in mesoscopic systems the speed of the computation is independent of the system size, but depends only on the kinetic parameters. Trade-offs involving time arise during the process of measuring the outcome of the computation. 
\par 
One could take the point of view that the measurement process is not an integral part of the computation, and that it is therefore unreasonable to charge this cost to the computation account. Indeed, it is true that there are many stochastic processes that happen in the world where the state of the system is never determined. However, information processing is only useful to the cell when the result is somehow utilised by the cell. For instance, during chemotaxis a bacterium determines the external concentration of some molecular species and then alters the motion of its flagella. More generally, if any computation is to have an effect on the world at all, then its result must be known. A computation that just happens without the result ever being communicated is thus not a  meaningful computation at all.  Measurement, therefore, is crucial to computation and cannot be separated.  
\par
 Within our model, there are two parts to  the measurement process. \one  The cost of  recording the result of the computation and \two the separation/restoration cost. As expected, the  cost of recording the measurement does not scale with the size of the measurement system. There are no fundamental reasons to assume that large systems are necessary in order to store information. Indeed, every electronic computer uses microscopic magnetic systems in order to store bits on disc drives. At a first glance more surprising is that the time-cost stems from the  time required to separate the system from the measurement device.
\par
The model of separation we used is, of course, not biologically realistic. Nonetheless, it reflects a real biochemical effect. The separation cost is directly related to the work required to break the bond between the external protein and the receptor, which features prominently in various biological incarnations of what is essentially the same effect.   A well known example is the classical insight that   the  unbinding rate of the ligand from the receptors at the cell surface fundamentally limits the ability of cells to measure molecular concentrations in their environment. This is   the ultimate origin of the  Berg-Purcell limit \cite{purcell,bialek2008,keizu}. Also relevant in this context is \cite{mylimitedpaper}, where it was shown how in a biological control system  trade-offs arise  from the dynamics of binding and unbinding of external ligands.
\par
Further  trade-offs, beyond the ones described here,  arise because the system $\mathfrak{S}$ is subject to noise. This means that a single measurement of the system will only result in a probabilistic answer.      In  order to increase the confidence  that the correct answer was obtained,  repeated measurements need to be performed  (see for example  \cite{myinterfacepaper}). In terms of our model, each of these sampling events comes at a measurement cost.    Altogether, the accuracy of the biochemical computation can be increased in two ways: Either  the EDC is made bigger, and thus noise reduced, while the cost is increased; or, alternatively,  the EDC remains constant but  its  result is sampled more frequently.   This   increases the time and the  cost.


\section*{References}


\end{document}